\documentclass[twocolumn,floatfix]{aastex63}

\usepackage{amssymb}
\usepackage{amsmath}
\usepackage{hyperref}
\usepackage{verbatim}
\usepackage{float}

\usepackage{graphicx}

\begin{document}

\title{The Highest-Redshift Balmer Breaks as a Test of $\Lambda$CDM}

\correspondingauthor{Charles Steinhardt}
\email{steinhardt@nbi.ku.dk}

\author[0000-0003-3780-6801]{Charles L. Steinhardt}
\affiliation{Cosmic Dawn Center (DAWN)}
\affiliation{Niels Bohr Institute, University of Copenhagen, Jagtvej 120, DK-2100 Copenhagen \O}

\author[0000-0002-5460-6126]{Albert Sneppen}
\affiliation{Cosmic Dawn Center (DAWN)}
\affiliation{Niels Bohr Institute, University of Copenhagen, Jagtvej 120, DK-2100 Copenhagen \O}

\author[0009-0006-7165-3828]{Thorbj\o rn Clausen}
\affiliation{Cosmic Dawn Center (DAWN)}
\affiliation{Niels Bohr Institute, University of Copenhagen, Jagtvej 120, DK-2100 Copenhagen \O}

\author[0000-0003-1561-3814]{Harley Katz}
\affiliation{Sub-department of Astrophysics, University of Oxford, DWB, Keble Road, Oxford OX1 3RH, UK}

\author[0000-0002-1515-995X]{Martin P. Rey}
\affiliation{Sub-department of Astrophysics, University of Oxford, DWB, Keble Road, Oxford OX1 3RH, UK}

\author[0009-0007-4720-7438]{Jonas Stahlschmidt}
\affiliation{Cosmic Dawn Center (DAWN)}
\affiliation{Niels Bohr Institute, University of Copenhagen, Jagtvej 120, DK-2100 Copenhagen \O}


\begin{abstract}
Recent studies have reported tension between the presence of luminous, high-redshift galaxies and the halo mass functions predicted by standard cosmology.  Here, an improved test is proposed using the presence of high-redshift Balmer breaks to probe the formation of early $10^4 - 10^5 M_\odot$ baryonic minihalos.  Unlike previous tests, this does not depend upon the mass-to-light ratio, stellar initial mass function, or star-formation history, which are all weakly constrained at high redshift.  We show that the strongest Balmer breaks allowed at $z = 9$ using the simplest $\Lambda$CDM cosmological model have $D_{4000} \leq 1.26$ under idealized circumstances and $D_{4000} \leq 1.14$ including realistic feedback models.  Since current photometric template fitting to JWST sources infers the existence of stronger Balmer breaks out to $z \gtrsim 11$, upcoming spectroscopic followup will either demonstrate those templates are invalid at high redshift or imply new physics beyond `vanilla' $\Lambda$CDM.
\end{abstract}


\section{Introduction}
\label{sec:intro}

The standard cosmological paradigm and hierarchical merging predict the timing with which small, primordial density fluctuations under the influence of gravity produce massive, compact halos.  Subsequently, baryons within those halos form stars and other structures recognizable as galaxies.  Remarkably, recent studies have reported galaxies that are so massive, at such high redshifts, that they appear to have formed even before their halos could have had time to collapse \citep{Steinhardt2016,Labbe2023} under the current standard model comprised of a cosmological constant dark energy and cold, collisionless dark matter (here termed `vanilla' $\Lambda$CDM).  As a result, the presence of massive, early galaxies has been interpreted as evidence for new physics beyond vanilla $\Lambda$CDM \citep{Steinhardt2016,Behroozi2018,BoylanKolchin2023}.  

However, this interpretation is challenging for three key reasons.  First, observations directly find \emph{luminous} galaxies at high-redshift rather than \emph{massive} ones. Stellar masses inferred from photometry are thus highly sensitive to assumptions on the mass-to-light ratio, which depend on the star formation history (SFH; \citealt{Conroy2013}) and assumed stellar initial mass function (IMF; \citealt{Steinhardt2023}).  Second, the possibility of a varying stellar baryon fraction \citep{Finkelstein2015} will affect the inferred halo mass which is being compared to the measured stellar mass. Finally, once a halo has virialized, additional time is required for the baryons within that halo to form stars. Thus, the observed stellar masses should not be compared against halo formation at the observed redshift, but rather at some higher redshift. Determining which redshift is most relevant again depends upon the star formation history, which is typically one of the least-constrained properties in photometric template fitting, and even more so at high redshift \citep{Laigle2016,Davidzon2017,Iyer2017,Weaver2022}. These significant systematic uncertainties makes it difficult to provide robust evidence for new physics beyond $\Lambda$CDM.

In this work, we propose an alternative, more robust test relying on the detection of the specific Balmer break spectral feature at 3646 \AA. The Balmer break comes from hydrogen atoms that have been excited into an $n = 2$ quantum state, and therefore is strongest at specific combinations of temperature and density such as those commonly found in the photospheres of A stars. Thus, the preconditions for producing a strong Balmer break are more complex than merely the creation of stellar mass. Rather, they reflect the characteristic timescale of stellar mass assembly, as younger stars have to evolve off their main sequence for A-type stars to dominate the galaxy's light. In fact, Balmer-breaks are routinely used as indicators of old stellar populations in low-redshift surveys (e.g., \citealt{Mobasher2005,Dunlop2013,Shahidi2020}), and traditionally interpreted to track a period of star formation and mass buildup and a subsequent quiescent phase lasting a few hundred Myrs. Given that such timescales approach the Hubble time at $z\geq8$, we argue here that their spectroscopic detection at high-redshift would provide a significantly more robust test of structure formation within vanilla $\Lambda$CDM, whose interpretation does not require assumptions about the IMF and SFHs. 

The quest for such high-redshift ($z\geq8$) Balmer breaks is long-standing (e.g., \citealt{Hashimoto2018,Laporte2021, Laporte2022}) but is rapidly expanding, with first photometric results from JWST implying $z=9-11$ Balmer breaks (e.g. \citealt{Adams2023,Atek2023,Furtak2023}) that might hint towards new physics. So far, these candidates have lacked the spectroscopic confirmation of the Balmer break which is key to distinguish them from a Lyman break \citep{HovisAfflerbach2021} or from nebular emission producing photometric excesses mimicking a break\footnote{A handful, including \citet{Hashimoto2018}, have spectroscopic redshift confirmation but none has rest-frame optical spectroscopy that confirms the Balmer break.} \citep{Roberts-Borsani2020,Carnall2023}. Large-scale spectroscopic programs with JWST are nonetheless underway (e.g., JADES, Bunker et al., in prep.; CEERS, \citealt{Finkelstein2023}; GLASS, \citealt{Treu2022}), making a theoretical quantification of the earliest Balmer break expected from vanilla $\Lambda$CDM particularly timely.

Since a Balmer break requires a stellar population which is a few hundred Myr old, the timing can be broken up into three required steps, which we systematically explore in this paper.  In \S~\ref{sec:clumps}, the formation time is calculated for the first virialized clumps capable of turning into stars.  Once star formation begins, the minimum time necessary to turn off star formation, if only for a few hundred Myr, is estimated in \S~\ref{sec:quenching}.  Finally, the time that must elapse between this turn off and the appearance of a recognizable Balmer break is calculated in \S~\ref{sec:timedelay}.  In \S~\ref{sec:discussion}, these components are put together and it is shown that many current galactic spectral energy distribution models fit to existing photometry predict Balmer breaks which, if verified spectroscopically, would already point towards new physics.  

\section{Forming stars in the first minihalos}
\label{sec:clumps}

The first stage in producing a Balmer break is to assemble a deep enough gravitational potential well to trigger radiative gas cooling and produce the compact clumps of baryons that will subsequently produce stars.  Here, two approaches are taken and compared to estimate this formation time.  First, an analytical estimate (as originally presented in \citet{Haiman1997}) is updated to current cosmological parameters. There are major differences between the concordance cosmological parameters when the original calculation was performed in 1997 and at present, and the effects of each difference is evaluated.  Second, the result is then compared to the formation time of the first Pop III stars in numerical simulations.  

There are significant differences between the assembly of baryons and dark matter.  Although cold dark matter structures form via hierarchical merging, baryonic collapse is complicated by diffusion damping \citep{Silk1968} and gas pressure, which greatly reduce anisotropies on small scales. Relative velocities between the baryon and dark matter fluids can lead to an additional delay in baryon clumping \citep{Tseliakhovich2010}, which is neglected in this analysis. Including this effect would delay baryonic structure formation, star formation, and quenching even further, so the bound presented here is overly conservative.

As a result, the smallest baryonic fluctuations collapse slower than larger structures, and will not allow for accelerated construction via hierarchical merging.  Instead, there is a characteristic mass scale corresponding to the first clumps of baryons to complete their collapse, likely between $10^4M_\odot$ and $10^5M_\odot$ (`minihalos'; \citealt{Haiman1997}).  
Given these fairly large gas masses, Population III (Pop III) stars likely form soon after the collapse of minihalos (see also e.g. \citealt{popiiireview, Bromm2013} for reviews). The formation time of minihalos thus provide a lower bound for the age of the Universe at the birth of the first stars. 

The \citet{Haiman1997} collapse time calculation assumed a cosmology with $(\Omega_b,\Omega_m,\Omega_\Lambda,h,\sigma_8) = (0.05,1,0,0.50,0.67)$.  Thus, conversion to the \citet{Planck2020} parameters of $(\Omega_b,\Omega_m,\Omega_\Lambda,h,\sigma_8) = (0.049,0.315,0.685,0.674,0.811)$ produces three significant adjustments:
\begin{enumerate}
    \item {Planck found an approximately 20\% increase in $\sigma_8$, the linear amplitude of oscillation on scales 8$h^{-1}$Mpc, over the 1997 parameters.  The collapse redshift is approximately linear in $\sigma_8$, so this will produce a collapse at $20$\% higher redshift than in the previous result.}
    \item {The spherical collapse time is often described as proportional to $1/H$ \citep{Kolb1990}.  However, this arises from dependence on the mass of the overdensity (or, equivalently, the matter density $\rho_m \propto \Omega_mH_0^2$), since the collapse time $t_c \propto M^{-1/2}$.  Thus, although Planck finds $h = 0.674 \pm 0.005$\footnote{The matter density would be slightly (but for this analysis, negligibly) larger than calculated here using $h$ from local Cepheid and supernova observations \citep{Riess2022}.}, larger than the 1997 value of $h = 0.5$, combined with the introduction of dark energy and conclusion that $\Omega_m = 0.315$ instead of $\Omega_m = 1$, $\rho_m \propto \Omega_mh^2$ decreases.  $H(z\approx 1100)$ decreases by 27\%, producing a corresponding increase in collapse time.}    
    \item {The same collapse time corresponds to a higher redshift when mapping age and redshift with the current concordance model, compared to the 1997 values.}
\end{enumerate}

\begin{figure}
    \centering
    \includegraphics[width=0.45\textwidth]{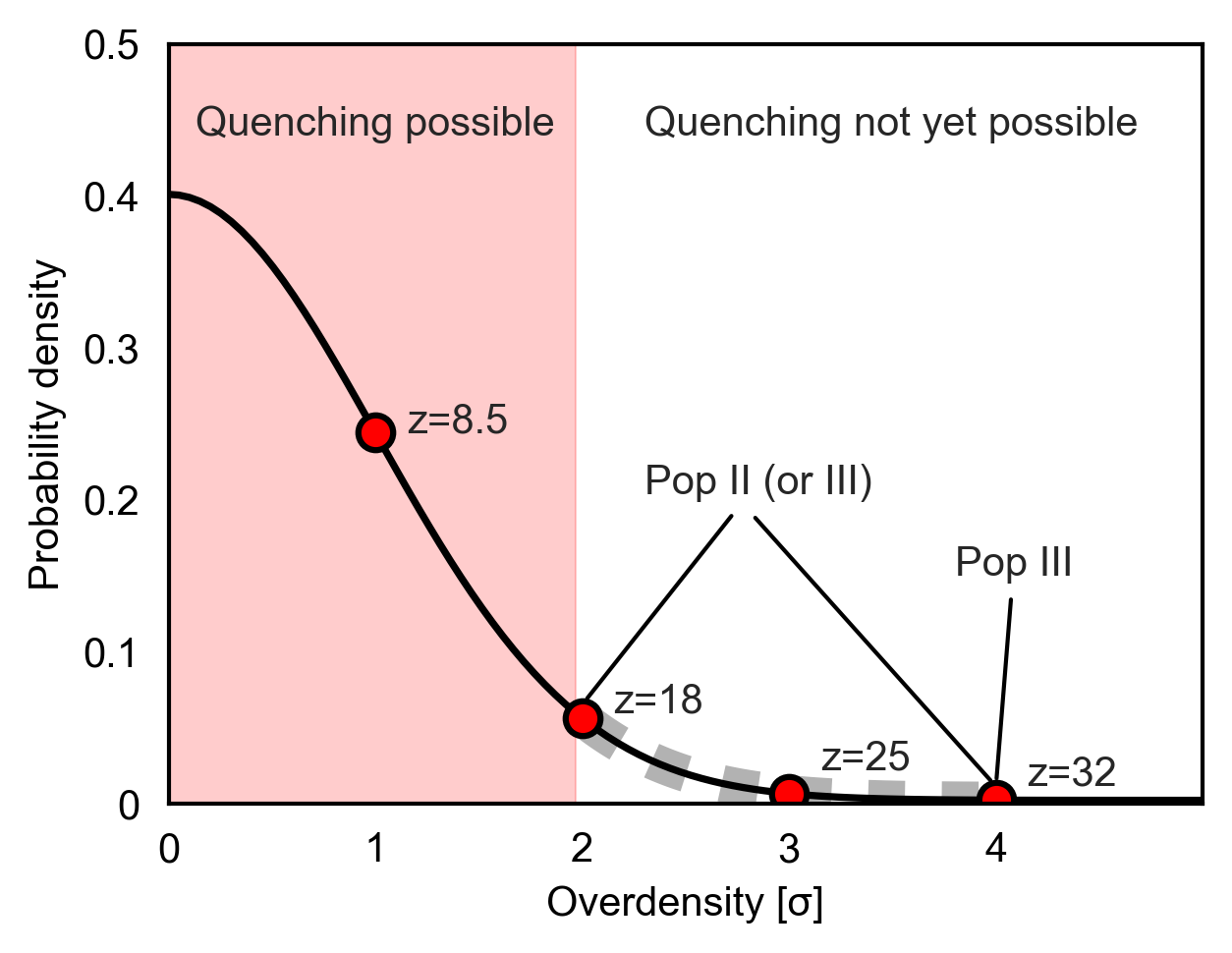}
    \caption{The distribution of overdensities that will form the first $10^4 - 10^5 M_\odot$ baryonic minihalos is well approximated as Gaussian.  The rare, most overdense regions (gray dashed) must form first Pop III and later Pop II stars.  Subsequent minihalos (red) can be heated by previous star formation and might not necessarily form stars at collapse.  Here, both semi-analytical arguments validated against simulations find the boundary to lie at $\sim 2\sigma$ overdensities collapsing around $z \sim 18$.  If a protogalaxy is discovered with a stellar population which quenched at $z > 18$, it would require new physics beyond `vanilla' $\Lambda$CDM.}
    \label{fig:overdensities}
\end{figure}

Taking these effects together, the \citet{Haiman1997} calculations can be updated to modern cosmology.  Since the first baryonic structures to collapse have masses of $10^4-10^5M_\odot$, between $\sim 10^4 - 10^6$ such clumps comprise a typical protogalaxy.  Thus, the most overdense of these will be $\sim 4-5\sigma$, and closer to $4\sigma$ for the earliest, and presumably smallest, protogalaxies \citep{MiraldaEscude2003}\footnote{The first structure to collapse in a large observed field will naturally occur earlier, but the time for that galaxy to quench (\S~\ref{sec:quenching}) will be comparable.}.  Therefore, early Pop III star formation is estimated to start at $z \sim 32$ under vanilla $\Lambda$CDM structure formation (Fig. \ref{fig:overdensities}). Slightly earlier than the $z \sim 27$ collapse-time estimated from the 1997 cosmology. 

This estimate of the earliest possible onset of star formation is in line with more realistic and modern cosmological simulations, with few reporting Pop III star formation above $z \approx 30-40$ (see e.g. \citealt{Jaacks2018, Liu2020} for recent works, and \citealt{Klessen2023} for a review). 

\section{Suppressing star formation for extended periods}
\label{sec:quenching}

A Balmer break requires the light of O and B stars to disappear and be replaced by A-type stars. Thus it is necessary to not only initially form stars but also to avoid subsequently forming newer, younger stars that would outshine the older population (or for differential dust attenuation to obscure those newer, younger stars but not the older stellar population; e.g. \citealt{Katz2019BB}). The precise stellar-evolution timescales associated with this aging process are investigated in Section~\ref{sec:timedelay}, but they are robustly $\geq 200$ Myr. Due to the lack of supermassive black holes at high-redshift, the prime candidate to suppress star formation is energy input from stars, first through their radiation and subsequently via supernovae explosions. 

Given the large uncertainties associated with modeling star formation and feedback at $z\geq10$, here a back-of-the-envelope estimate is provided for generating enough stellar mass and feedback energy to quench star formation for several hundred million years. Then star formation histories are quantitatively examined with the state of the art radiation hydrodynamical \textsc{sphinx} simulation (\citealt{Rosdahl2018, Rosdahl2022}).

In order to quench star formation, enough gas needs to be transformed into stars for them to significantly heat their surroundings and balance the very quick cooling times of the early Universe \citep[e.g.][]{White1978,Dekel1986}. Most of the first minihalos form in isolation, in density peaks, and the short main sequence lifetimes associated with massive Pop~III stars make them inefficient as a sustained heating source. Consequently, one can speculate that efficient quenching requires the larger over-densities of the Universe to collapse and provide the necessary stellar feedback (via a Pop~II stellar population). A $2\sigma$ overdensity collapses at $z = 18$ (Fig. \ref{fig:overdensities}), which as below is a good estimate for when gaps can begin appearing in the star formation histories of high-redshift dwarf galaxies.

To estimate this more quantitatively, the star formation histories in Version 1 of the SPHINX Public Data Release (SPDR1) are used. \textsc{sphinx}$^{20}$  models the cosmological formation of the first galaxies in an (20 comoving Mpc)$^3$ `average' volume (i.e. with particularly over-dense regions favoring earlier collapse) with enough numerical resolution (10 pc spatial resolution and $2 \times 10^5\, M_{\odot}$ dark matter particle mass) to start resolving the interstellar medium. The formation history is extracted from all stars within galaxies at $z=10$, 9, \& 8 that have ${\rm SFR}\geq 0.3 \,{\rm M_{\odot}}\, \text{yr}^{-1}$ to mimic a flux-limited survey (see \citealt{Choustikov2023} for further details), and systematically searched for 100, 200, 250 Myr gaps in star formation rates. 

\begin{figure}
    \centering
    \includegraphics[width=0.45\textwidth]{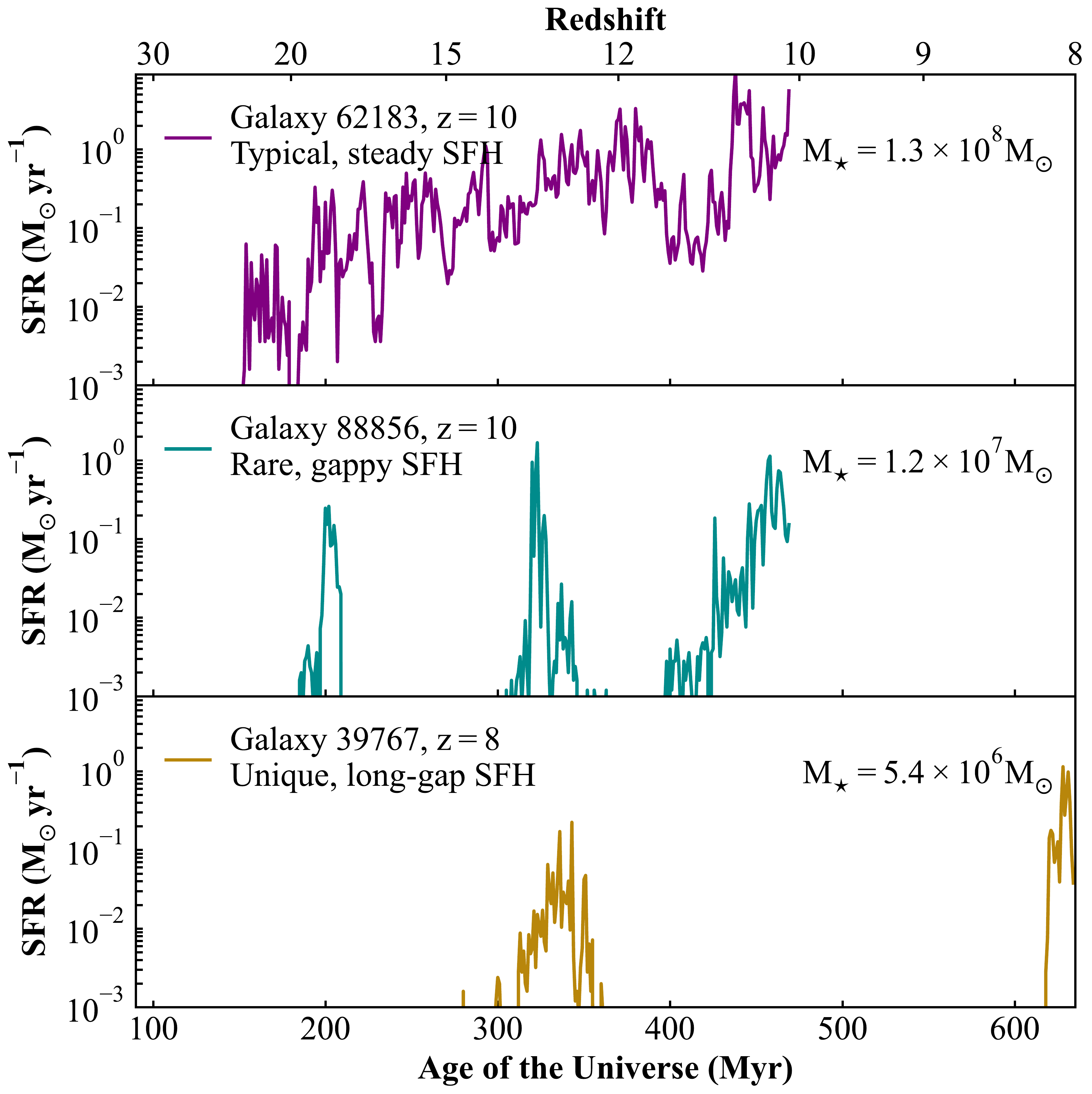}
    \caption{Sample star-formation histories of high-redshift galaxies from data products of the \textsc{sphinx} simulation (\citealt{Rosdahl2018, Rosdahl2022, Choustikov2023}). Rising, continuous star formation histories are the most common occurrence (top) at z=8, 9 and 10, but rare examples show extended gaps induced by stellar feedback (middle panels) in lower-mass objects. One, unique case of a very low-mass dwarf showcasing a $\geq 250$ Myr gap is found (bottom), comparable to the stellar-evolution timescales necessary to produce strong Balmer breaks (Figure~\ref{fig:spectra}).}
    \label{fig:sphinx}
\end{figure}

Selected examples of such star formation histories are shown in Figure~\ref{fig:sphinx}. Nearly all star formation histories are continuous and rising (e.g. top panel), with several examples exhibiting star formation at $z\geq20$ and validating the analytical timescales from \S~\ref{sec:clumps}. Four galaxies are found at $z=10$ with $\geq 100$ Myr gaps (middle panel), but none with longer. The bottom panel shows the only example of a low-mass, $z=8$ galaxy exhibiting a $\geq 250$ Myr gap, approaching the timescale necessary to create a strong Balmer break in the absence of dust (\S~\ref{sec:timedelay}). 

The lack of $\geq 250$ Myr gaps at very high redshift ($z\geq 12$) in the \textsc{sphinx} data suggests that of the star-forming populations that are likely observable with \emph{JWST}, very few are expected to exhibit strong Balmer breaks (see also \citealt{Binggeli2019}). Nonetheless, a key learning from inspecting these star formation histories is that extended gaps due to stellar feedback can be generated as early as $z=18$, even if the details of their lengths depend on the specific model and astrophysical conditions. In the remaining of this paper, it is thus assumed that quenching star formation for 300 Myr at $z\approx18$ is within the realm and uncertainties of current models, and the following section quantifies the time needed for stars formed at that time to evolve off the main sequence and create strong Balmer breaks.

\section{From Quenching to Production of the Balmer Break}
\label{sec:timedelay}

Finally, once a galaxy has quenched, enough time must elapse for the most massive stars to evolve off the main-sequence so that the observed spectral energy distribution is dominated by stars with cool enough photospheres to produce a Balmer break. To gain a clean intuition of the timescales involved and how they relate to the characteristics of the Balmer break, a grid of models was run using the Flexible Stellar Population Synthesis (FSPS; \citealt{Conroy2009b}) library to model the aging of a single stellar population (SSP). Using an SSP is conservative as any more complex star formation history will take longer to produce a Balmer break. Here, a fiducial model with $[Z/H] = -1.7$, $A_V = 0.1$, and an IMF corresponding to a gas temperature of 60~K is chosen. As described in the remainder of this section, a comparison with previous work shows that the primary results here are not strongly sensitive either to the specific parameters or to the choice of the FSPS library.

\begin{figure}[t]
    \centering
    \includegraphics[width=0.46\textwidth]{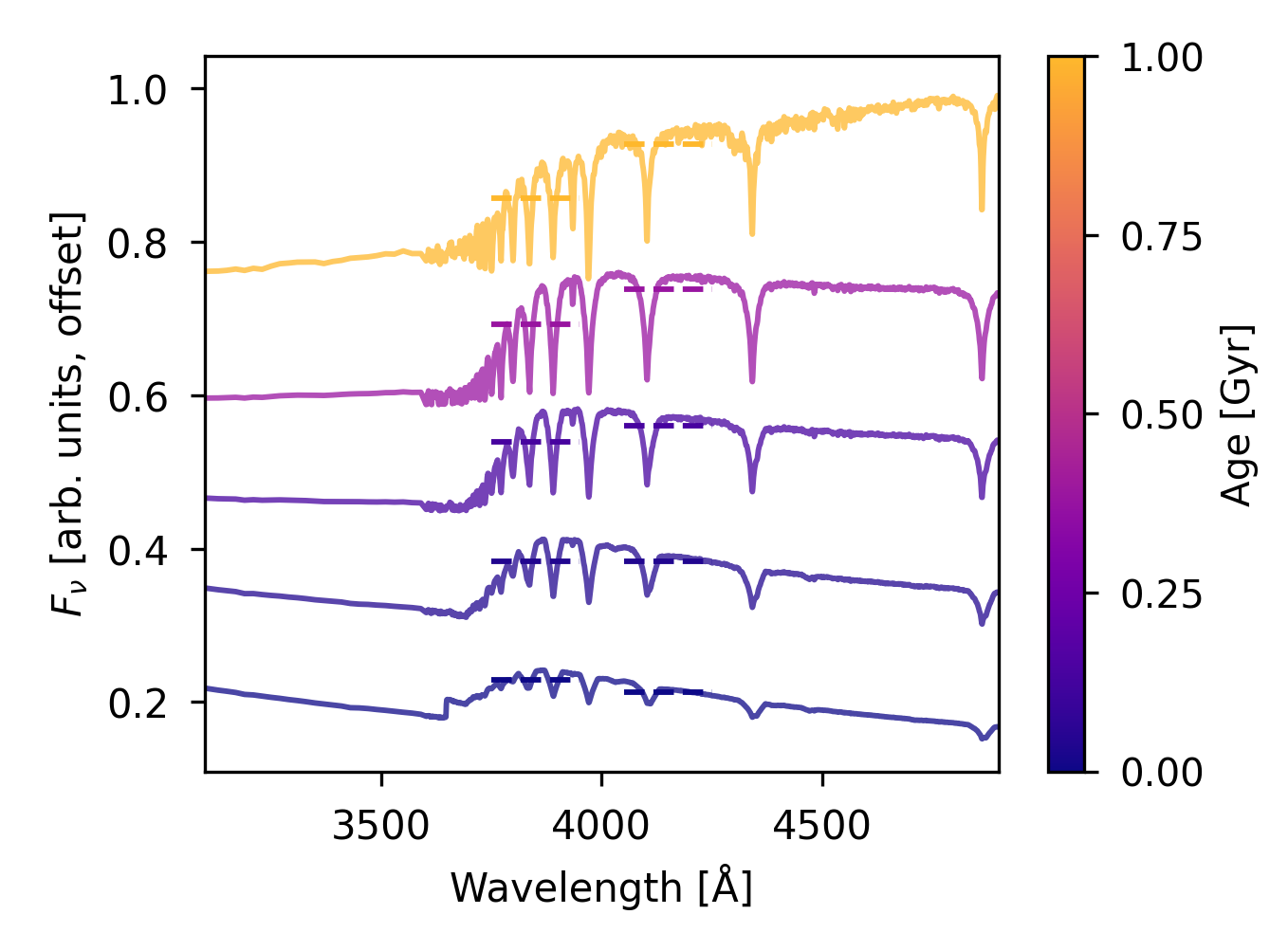}
    \caption{Spectral energy distributions around 4000\AA~for simulated galaxies composed of a single stellar population of Pop II stars with ages 25, 63, 158, 398, and 1000 Myr.  The depth of the Balmer break grows rapidly with increased age to $\sim 300$ Myr and more gradually after that.  This strength of this feature is often described in terms of $D_{4000}$, the slope between the two regions indicated as dashed lines.  $D_{4000}$ is negative for very young populations and grows as the population ages towards 1 Gyr.}
    \label{fig:spectra}
\end{figure}

As the age of the stellar population increases, the spectral energy distribution begins to exhibit a visible Balmer break (Fig. \ref{fig:spectra}).  For the remainder of this work,
\begin{equation}
    D_{4000} = \frac{\int_{4050\overset{\circ}{A}}^{4250\overset{\circ}{A}} F_{\nu} \ d\lambda}{\int_{3750\overset{\circ}{A}}^{3950\overset{\circ}{A}} F_{\nu} \ d\lambda}
\end{equation}
is used a standard measure of the depth of the break \citep{Bruzual1983,Poggianti1997}.

\begin{figure*}[ht]
    \centering
    \includegraphics[width=0.95\textwidth]{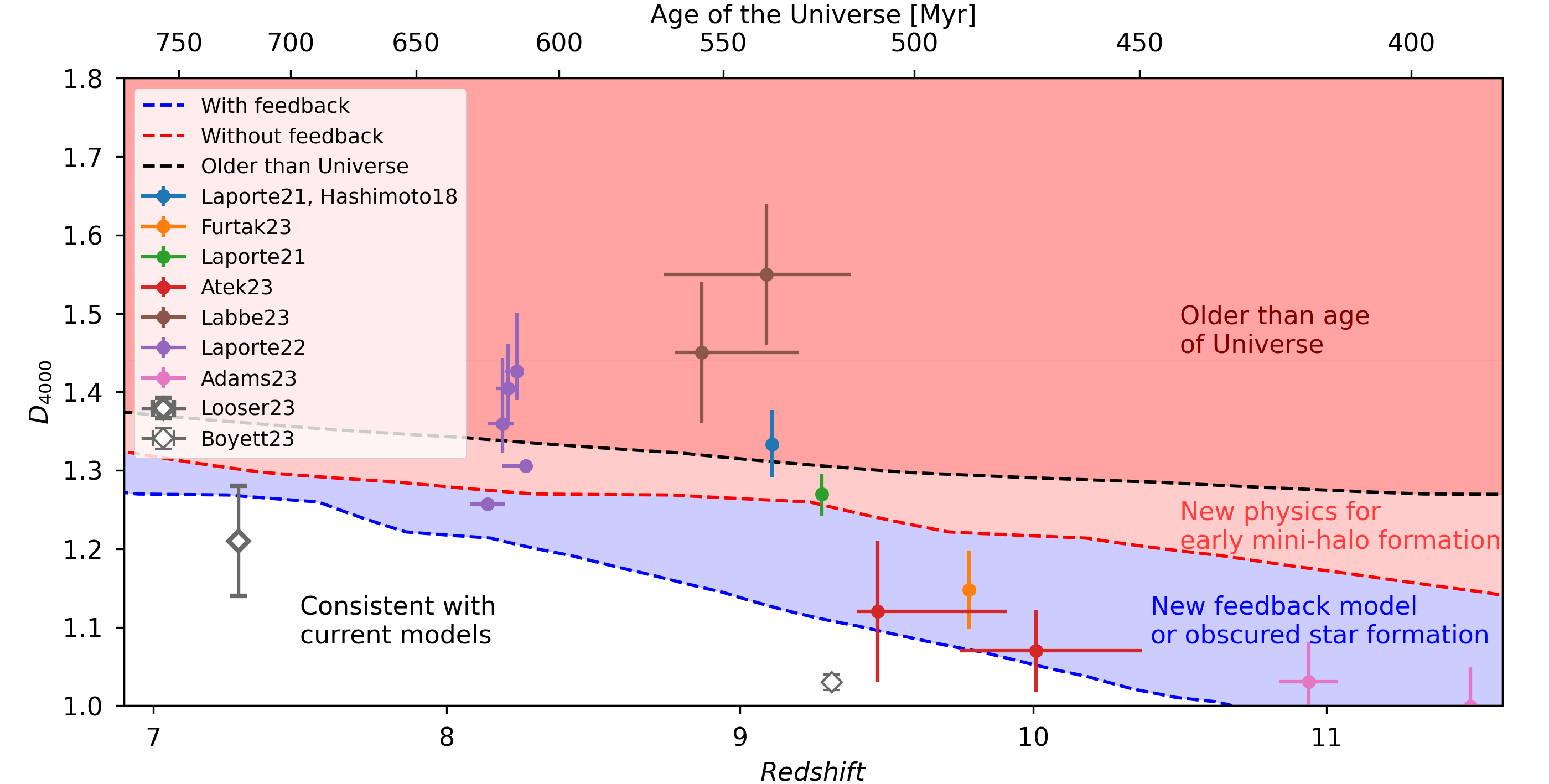}
    \caption{A selection of galaxies with photometrically-selected Balmer breaks \citep{Adams2023,Atek2023,Furtak2023,Hashimoto2018,Labbe2023,Laporte2021,Laporte2022}, as reported by the original authors (MACSJ1149-JD1 is first discussed in \citet{Hashimoto2018} with updated parameters taken from \citet{Laporte2021}).  The maximum possible $D_{4000}(z)$ as calculated here is shown for comparison. Taken at face value, the strongest, highest-redshift Balmer breaks predicted by current photometric studies are inconsistent with `vanilla' $\Lambda$CDM and require new physics. In fact, the inferred stellar population would be substantially older than the age of the Universe for current cosmological parameters and require a far more exotic model. Weaker photometric breaks could be reconciled by alterations to feedback models, obscured star formation or new physics driving earlier mini-halo collapse and structure formation. Spectroscopic confirmation is thus paramount to confirm the nature of these early Balmer breaks, with the current highest-redshift spectroscopically-confirmed breaks \citep{Looser2023,Boyett2023} being consistent with current models.}
    \label{fig:d4000}
\end{figure*}

It should be emphasized that two other common definitions of the depth of the break exist.  For the fiducial 300 Myr SSP used here (Fig. \ref{fig:d4000}, red dashed line), the \citet{Bruzual1983} $F_{\nu,4150}/F_{\nu,3850}$ is 1.24.  For comparison, an alternative definition of $D_{4000} = F_{\nu,4200}/F_{\nu,3500}$ \citep{Binggeli2019} as used in BPASSv2.2.1 \citep{Stanway2018} for the same population yields $D_{4000} = 2.43$, and the narrow-band $D_n(4000) = 1.11$ \citep{Balogh1999}, so it is critical to note which definition is used when comparing against observations.

\begin{figure*}
    \centering
    \includegraphics[width=0.95\textwidth]{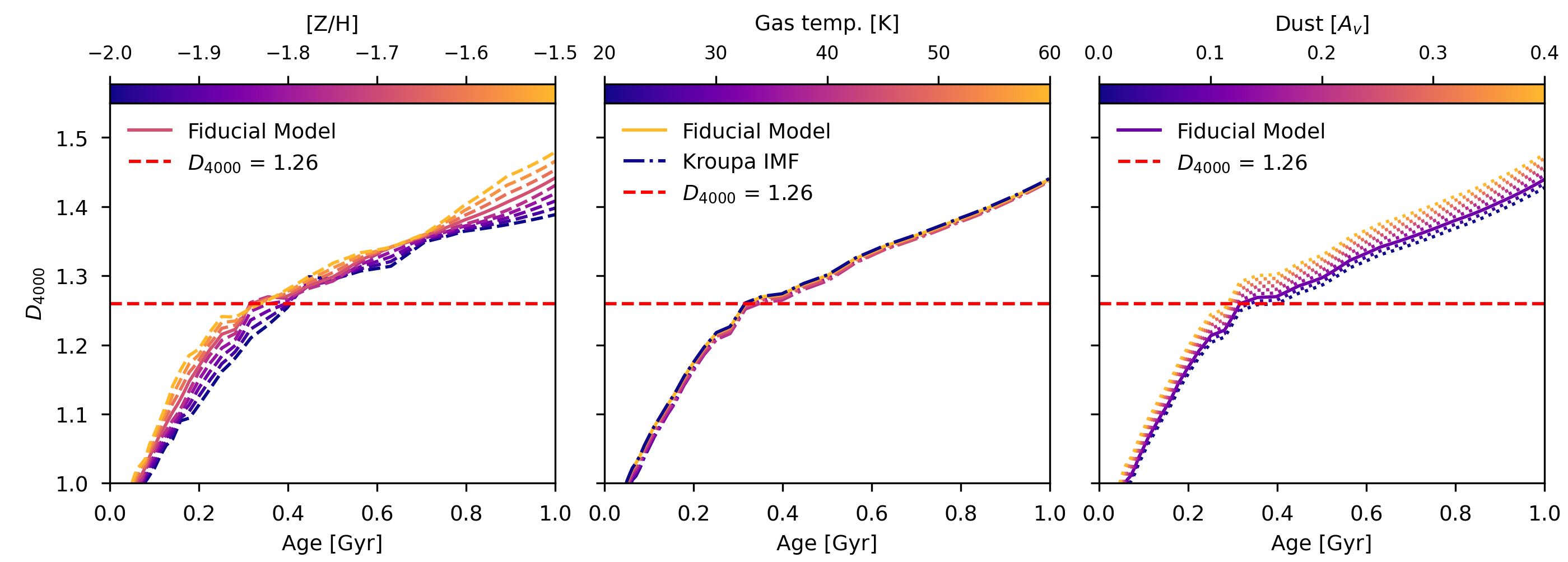}
    \caption{Growth of $D_{4000}$ measuring the strength of the Balmer break as a function of stellar population age, varying metallicity (left), the IMF (middle) and dust extinction (right). For the fiducial model at $[Z/H]= -1.7, \ A_V = 0.1,$ and a IMF corresponding to a gas temperature of 60 K, the Balmer break grows rapidly with age until $D_{4000} \approx 1.26$, which is followed by a more gradual growth.  Here, a spectrum with $D_{4000} > 1.26$ (red, dashed line), such as those reported via photometric template fitting to some ultra-high redshift JWST sources, is described as having a prominent Balmer break.  The timing of the earliest possible Balmer break is weakly sensitive to assumptions of the IMF (middle) and extinction (right), but has a stronger dependence on metallicity (left).}
    \label{fig:paramscomp}
\end{figure*}

The evolution of this measure as a function of age in is shown in Fig. \ref{fig:paramscomp}. For very young stellar populations, $D_{4000} < 1$ because of the blue continuum produced by the most massive stars.  As the stellar population ages and young O and B stars evolve off the main sequence quickly, $D_{4000}$ grows rapidly. This growth then becomes more gradual with time, as the last lower-mass B stars already exhibit weaker Balmer breaks and are slowly evolving off the main sequence. Galaxies in this latter phase are typically described as exhibiting Balmer breaks.  The breakpoint in $D_{4000}$ as a function of age lies near $D_{4000} = 1.26$ (Fig. \ref{fig:paramscomp}, which is weaker than the Balmer breaks reported using photometric template fitting to \emph{Hubble} and \emph{JWST} data (\S~\ref{subsec:templatefits}).

In order to model populations typical of those that could produce the earliest Balmer breaks, the following assumptions are tested:
\begin{itemize}
    \item {As metallicity decreases, stellar photospheres increase in temperature at fixed mass.  As a result, the Balmer break is produced by lower-mass stars than at higher metallicity, so the stellar population must be older for these lower-mass stars to dominate the spectrum and produce a Balmer break.  Since the time required to produce a Balmer break is longer towards lower metallicity (Fig. \ref{fig:paramscomp}, left), here $[Z/H] = -1.7$ is chosen in an attempt to provide a conservative lower bound on the time required to produce a Balmer break.  Although there is some justification for such a high Pop II metallicity based on Galactic Pop~II stars \citep{Baade1944,Rix2022}, this value is almost certainly an overestimate of the metallicity of very early Pop~II stars.  }
    \item {Although dust can contribute to an earlier production of a 4000 \AA~break, there is likely not enough time for dust to be produced in significant quantities by $z \sim 20$ \citep{Gall2011,Ferrara2016}.  Here, a maximum extinction of $A_V = 0.1$ mag is taken as a conservative upper bound on high-redshift dust.  Photometric template fitting finds that $A_V < 0.1$ is more likely even at $z = 10$ \citep{Steinhardt2023}.  Regardless, the time-scale for producing a Balmer break is relatively insensitive to $A_V$ (Fig. \ref{fig:paramscomp}, middle) as long as the dust is evenly distributed.}
    \item {Since the cosmic microwave background temperature at $z = 20$ is slightly under 60 K, a bottom-lighter stellar initial mass function than in the Milky Way is very likely required \citep{LyndenBell1976,Larson1985,Sneppen2022,Steinhardt2023}.  A reduction in metallicity, as is likely at very high redshift, will act in the same direction.  Here, the 60 K IMF prescription from \citet{Jermyn2018} is assumed for these first Pop II stars.  Nevertheless, the timing of Balmer break formation is only very weakly dependent on the IMF, since the most massive stars must nearly all die regardless of their exact relative fractions of the stellar population (Fig. \ref{fig:paramscomp}, right).}
\end{itemize}

From this systematic exploration, it is concluded that 340 Myr are necessary to produce a Balmer break with $D_{4000} > 1.26$ after star formation. Combined with the timescales for initial collapse and subsequent quenching, the earliest possible Balmer break with $D=1.26$ cannot occur until the Universe is approximately 550~Myr old (i.e. $z=8.9$). The maximum possible $D_{4000}$ as a function of redshift is shown in Fig. \ref{fig:d4000}.

\section{Discussion}
\label{sec:discussion}

The timing of the formation of the first stars and galaxies is intrinsically linked to structure formation within our cosmological model. Despite the difficulties of modeling such astrophysical processes, we have shown that strong Balmer breaks at high-redshift can provide a test of early structure formation and potentially new physics beyond vanilla $\Lambda$CDM.  

We find that a conservative upper limit for the highest-redshift Balmer break is $z \sim 9$, assuming a single burst of self-quenching very early star formation. Leveraging the state-of-the-art \textsc{SPHINX} simulation of early galaxy formation and reionization, we find that a more realistic estimate is closer to $z = 7.5$ given our current understanding of star formation and feedback in the first galaxies in an average-density environment of the Universe. 

Interestingly, photometric observations of Balmer-breaks in this regime have already been reported from photometric template fitting. This includes objects with $z_{\rm phot} \approx 7-9$ \citep{Roberts-Borsani2020,Laporte2022,Labbe2023}, $z_{\rm phot} \approx 9-10$ \citep{Hashimoto2018,Laporte2021,Furtak2023,Atek2023}, and even a couple of candidates at $z_{\rm phot} \approx 11 $ \citep{Adams2023}.  If taken at face value\footnote{Redshifts and uncertainties shown in Fig. \ref{fig:d4000} are those reported by the original authors, even where the reported uncertainty is likely far smaller than truly attainable for photometric redshifts.  $D_{4000}$ was provided by the original authors for \citet{Adams2023,Boyett2023,Hashimoto2018,Furtak2023,Laporte2021,Laporte2022,Looser2023} and estimated from figures for the remainder.}, these are already hinting at new physics (see Fig. \ref{fig:d4000})

However, there is an absolute necessity for spectroscopic confirmation of these breaks to draw robust conclusions.  Indeed, both initial NIRSpec spectroscopy \citep{Carnall2023} and template fits using physics tuned to ultra-high redshift galaxies \citep{Steinhardt2023} suggest these features may instead be due to strong nebular emission.  

Conversely, a very weak Balmer break has already been inferred from spectroscopy at $z=9.3127\pm 0.0002$ \citep{Boyett2023}.  Similarly, the \citet{Hashimoto2018} object has spectroscopic confirmation of $z = 9.1096 \pm 0.0006$ from an 88 $\mu$m oxygen line, but the Balmer break is only detected photometrically. 

One mechanism suggested for producing such an early Balmer break would be a combination of an older stellar population and dust-enshrouded younger population. This might allow a Balmer break to be produced even as a galaxy continues to form stars \citep{Katz2019BB}.  Such a mechanism would mimic the appearance of quenching, and thus allow Balmer breaks in the blue region in Fig. \ref{fig:d4000}, where no new physics is required. Such a model was successfully used to explain MACS1149\_JD1. However, most of the photometric Balmer breaks that have been reported would be deep enough to require star formation earlier than expected for vanilla $\Lambda$CDM, which means if verified, they could not be explained by differential dust alone.

\subsection{Are ``Impossible'' Photometric Balmer Breaks from JWST Genuine?}
\label{subsec:templatefits}

Previous studies have also found that the apparent stellar masses of these high-redshift \emph{Hubble} and \emph{JWST} galaxies are too high to have been produced by cold, collisionless dark matter \citep{Steinhardt2016,Labbe2023}.  This tension becomes far stronger in combination with a deep presumed Balmer break, since producing the earliest possible Balmer breaks requires an older stellar population (\S~\ref{sec:timedelay}) and likely a small stellar baryon fraction (\S~\ref{sec:quenching}), both of which greatly reduce the halo mass-to-light ratio.  Models which constrain ultra-high redshift objects to have physically-plausible stellar initial mass functions and extinction instead produce far lower masses and younger stellar populations, consistent with vanilla $\Lambda$CDM \citep{Steinhardt2023}.  These same templates also predict that the red photometric ``bumps'' are produced by strong nebular emission lines rather than a Balmer break, potentially resolving both problems via the same effect. 

This explanation also predicts emission lines must be stronger than seen at lower redshifts, but comparable to those seen in early \emph{JWST} spectroscopy at $z > 6$ \citep{Carnall2023,Cameron2023,Sanders2023}.  In that way, upcoming spectroscopic followup of these early \emph{JWST} candidates will not only determine which model is the better fit, but also simultaneously determine whether current cosmological models pass or fail both the stellar mass and Balmer break tests.

The feature observed in the reddest bands (which is commonly interpreted as a Balmer break) has even been proposed as a selection mechanism in an attempt to confirm the ultra-high redshifts of targets which would be difficult to pinpoint from a Lyman break alone \citep{Labbe2023}. However, as shown in this work, for any of these to truly be strong Balmer breaks would be inconsistent with vanilla $\Lambda$CDM. Regardless, such a selection mechanism will still work if the red photometric bumps are due to strong nebular emission lines, although it would imply the current inferred photometric redshifts are likely to be slightly overestimated. For example, stronger [O\textsc{iii}]+H$\beta$ emission lines would produce a red bump in photometry at 15-20\% longer wavelengths than the Balmer-break feature, resulting in correspondingly lower redshifts.  Finally, it is also possible that some of these photometric redshifts will turn out to have been catastrophic overestimates, as was already the case for some early \emph{JWST} sources (e.g., \citealt{ArrabalHaro2023}).

\subsection{Cosmological Models for $z > 9$ Balmer Breaks}

At present, it would seem that the most likely outcome is that spectroscopic followup will not discover Balmer breaks in contradiction with vanilla $\Lambda$CDM, but rather that the spectral energy distributions predicted from lower-redshift analogues are not representative of the first galaxies.  However, given the possibility that some of these galaxies will indeed exhibit ultra-high redshift Balmer breaks, it is worth considering what modifications to vanilla $\Lambda$CDM would be needed to produce them.  

Two possible solutions include modifying the IMF so that massive stars never form and the Balmer break can form earlier in the evolution of the SSP, or modifying the matter power spectrum (e.g. via a modification to dark matter or other process) to allow for earlier structure formation. The former is a poor physical solution as due to increased heating from the CMB and other sources, the high-redshift IMF is expected to be bottom-light rather than bottom-heavy \citep{Jermyn2018}. 

Modifying dark matter interactions will also impact structure formation. Warm dark matter models typically inhibit structure formation \citep{Murray2013,Schneider2013,Pacucci2013}, so a CDM scenario is still most likely. Self-interacting cold dark matter models \citep{Spergel2000,Tulin2018} may be a more promising approach for earlier structure formation. Baryon-dark matter interactions have also been proposed \citep{Barkan2018DMBarInteractions} to hasten gas cooling and star formation in an attempt to explain possible detection of the global 21~cm absorption signal \citep{Bowman2018} (but see also \citealt{Barkana2018NotDMBarInteractions}).  

Primordial magnetic fields \citep{Turner1988,Ratra1992} are known to enhance power at high {\it k}-modes of the matter power spectrum due to the Lorentz force \citep[e.g.,][]{Wasserman1978,Kim1996}. This can lead to structure formation earlier than vanilla $\Lambda$CDM, although measurements/upper limits of the high-redshift electron optical depth \citep[e.g.,][]{Heinrich2021,Planck2020} constrain the amount of possible early structure formation \citep{katz2021} due to its impact on H{\small I} reionization.

Another possibility would be a model that produces primordial structure.  Primordial black holes have been proposed via several mechanisms \citep{Carr2020,VillanuevaDomingo2021}.  These would then seed more rapid growth of the first baryonic minihalos, and thus the first stars.  

The most extreme photometric Balmer breaks, if verified, would require a stellar population which is significantly older than the age of the Universe using the \citet{Planck2020} cosmology (Fig. \ref{fig:d4000}).  Such a population could not be produced even by primordial structure formation or dark matter modifications, and would instead be entirely incompatible with $\Lambda$CDM models.  Given the overwhelming observational evidence for the $\Lambda$CDM cosmological paradigm (cf. \citet{Bull2016}), this would seem to be the least likely outcome.

Fortunately, the answer to this question is likely to be provided in the very near future.  \emph{JWST/NIRSpec} followup of several photometrically-selected $z > 9$ candidates with red bumps has already been performed, with the results currently in preparation.  If even a single one of these sources indeed exhibits a clear Balmer break with sufficiently high $D_{4000}$, as predicted by many photometric template fits, then vanilla $\Lambda$CDM will be ruled out.  However, if they all instead caused by strong nebular emission, modifications in templates for high redshift along the lines of those proposed for bottom-light IMFs \citep{Sneppen2022,Steinhardt2023} will be required instead. \newline 

The authors would like to thank Sebastien Aagaard, Kit Boyett, Lukas Furtak, Zoltan Haiman, Thomas Harvey, Troels Haugb\o lle, Anne Hutter, Nicolas Laporte, Tobias Looser, Bahram Mobasher, Vadim Rusakov, Luka Vujeva, and Darach Watson for helpful comments and discussions.  The Cosmic Dawn Center (DAWN) is funded by the Danish National Research Foundation under grant No. 140. MR is supported by the Beecroft Fellowship funded by Adrian Beecroft.

\bibliographystyle{aasjournal}
\bibliography{refs.bib} 



\label{lastpage}
\end{document}